\documentclass{elsart}
\usepackage{epsfig}
\usepackage{amssymb}

\begin{document}
\begin{frontmatter}

\title{On the non-Boltzmannian nature of quasi-stationary states in long-range interacting systems}

\author
{Constantino Tsallis}  \ead{tsallis@cbpf.br}
\address {Centro Brasileiro de Pesquisas Fisicas, Rua Xavier Sigaud 150, 22290-180 Rio de Janeiro-RJ, Brazil }

\author
{Andrea Rapisarda} \ead{andrea.rapisarda@ct.infn.it}
\address{Dipartimento di Fisica e Astronomia,  Universit\'a di Catania,\\
and INFN sezione di Catania,  Via S. Sofia 64, I-95123 Catania, Italy}

\author
{Alessandro Pluchino} \ead{alessandro.pluchino@ct.infn.it}
\address{Dipartimento di Fisica e Astronomia,  Universit\'a di Catania,\\
and INFN sezione di Catania,  Via S. Sofia 64, I-95123 Catania, Italy}

\author
{Ernesto P. Borges} \ead{ernesto@ufba.br }
\address { Escola Politecnica, Universidade Federal da Bahia, Rua Aristides Novis 2, 40210-630 Salvador-BA, Brazil }

%
%

\begin{abstract}


We discuss  the non-Boltzmannian   nature  of quasi-stationary states in the  Hamiltonian Mean Field (HMF) model, a paradigmatic  model  for long-range interacting classical many-body systems. We present a theorem excluding the Boltzmann-Gibbs exponential weight in Gibbs $\Gamma$-space of microscopic configurations, and comment  a  paper  recently published by Baldovin and Orlandini (2006). On the  basis  of the  points here   discussed,   the ongoing debate on the  possible application, within appropriate limits, of the generalized $q$-statistics  to long-range Hamiltonian   systems   remains open.

\end{abstract}

\end{frontmatter}

\section{Introduction}
The  Hamiltonian  Mean Field (HMF) model is  a  paradigmatic \textit{toy model}   for  long-range  interacting  Hamiltonian   systems \cite{hmf}.
In a recent paper \cite{fulvio2},
Baldovin and Orlandini  presented new molecular dynamics numerical results for the HMF model  in contact with a thermal bath. They conclude that the energy distribution in its  {\it quasi-stationary state} (QSS) is of the Boltzmann-Gibbs (BG) form. In view of their own numerical results, we cannot agree with their conclusions. In this paper we present  three points that are relevant for the interesting questions raised in \cite{fulvio2}.

\begin{figure}
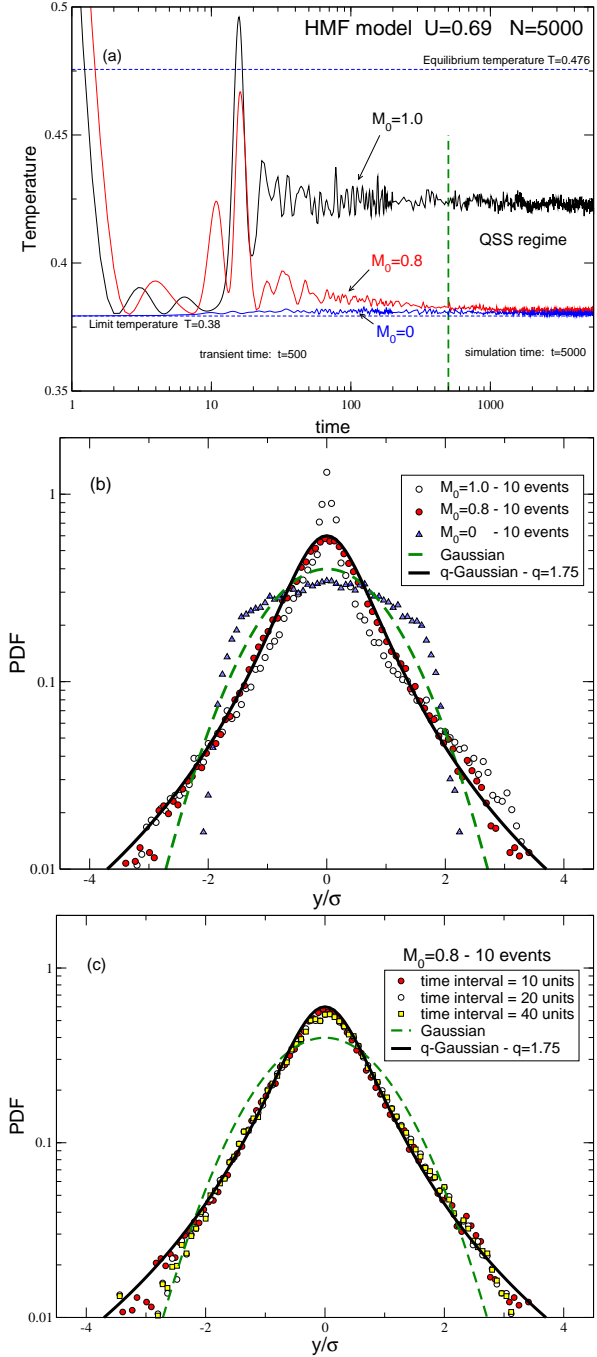

\begin{center}
\epsfig{figure=fig1-a.eps,width=7.8cm,angle=0}
\epsfig{figure=fig1-b.eps,width=7.8cm,angle=0}
\epsfig{figure=fig1-c.eps,width=7.8cm,angle=0}
\end{center}
\caption{ (a)
The temperature as a function of time  for  the  HMF model at $U=0.69$ and $N=5000$ obtained for three  initial magnetization values, respectively $M_0=1,\,0.8,\,0$, and averaged over $10$ events. The system remains trapped  in a QSS at a temperature value  lower  than the   equilibrium one. The latter is drawn together with the  limit temperature at this  energy \cite{epn-rap}. (b) The  pdf  of   stochastic  values $y$ obtained  by summing rotators momenta in the   QSS regime  illustrated  in  panel (a), see text. Different initial magnetizations $M_0$ are  reported. A Gaussian (dashed curve) and a  $q$-Gaussian with $q=1.75$ (full curve) are also reported for comparison. (c) For the initial magnetization $M_0=0.8$, we plot the pdfs obtained changing the  time interval for  recording the   momenta of the rotators. No significant change is observed. See text for further details.
}
\end{figure}

\section{Discussion}

In what  follows  we  present three important  points regarding  the  non-Boltz\-mannian nature  of the QSS  in the HMF  model.\\

\emph{(i) The nature of the QSS in the canonical setup presents differences with that in the microcanonical setup.}\\

It has already been demonstrated that, in the microcanonical ensemble,  the anomalous dynamics (concerning the  presence  of a glassy phase, hierarchical structures, velocity correlations, aging, vanishing Lyapunov exponents, etc),
and therefore the statistical description of the QSS, is strongly dependent on the initial conditions \cite{epn-rap,liap,chavanis}. \textit{A fortiori}, it is thus not at all justified to {\it extrapolate}  the conclusions for the canonical QSS studied in \cite{fulvio2,fulvio1} to the
microcanonical ones. Although the authors  themselves   state in \cite{fulvio1} that the canonical QSS's are only \textit{reminiscent} of the
microcanonical ones (in fact they differ in the lifetimes, the relaxation times, and, most probably, also in the correlations), they neglect it in \cite{fulvio2}.
In order   to  stress  this point, in Fig. 1 (a)    we  plot  three numerical simulations  in the  microcanonical ensemble  (averaged over $10$ events) for $U=0.69$, $N=5000$ and for different  initial magnetizations $M_0$. Water-bag initial conditions were used \cite{epn-rap}. Following the Central Limit Theorem aspects recently revealed in \cite{UCC}, we  illustrate 
in Fig. 1 (b) the non-Boltzmannian   nature  of this  QSS  regime   by  studying the  
  probability  density functions (pdfs) of the  stochastic  values $ y$ defined as follows: $ y_j=\sum_{i=1}^n  l_j(i)  $, for  $j=1,2,...,L$, where  $l_j(i) $
are the   momenta of the $jth$-rotator taken at fixed intervals of time $i$ inside the QSS   regime. We considered $L=5000$ and  n=500  intervals, one  every $10$ units of time. The length of the time interval was varied as shown in Fig. 1 (c) for the initial magnetization $M_0=0.8$ without any relevant change in the pdfs. An average over 10 events was also considered, in order to have $50000$ values  for the  $y$ variables for each  pdf. In case  of no global correlations, as one would expect at equilibrium, the resulting  pdf  should be a Gaussian. As  Fig. 1(b) shows, in our case, the Gaussian distribution does {\it not} occur. In fact we have   pdfs whose shape  depends  on the  initial magnetization $M_0$ and the central region (up to a few mean standard deviations) of most of them is  well fitted by $q$-Gaussians.  In particular   for $M_0=0.8$  we get a $q$-Gaussian with $q=1.75$  as  in  \cite{UCC}.   We do not think that this is a universal     value in this  case, since  we do not  have  a strictly zero largest Lyapunov
exponent as in \cite{UCC}. For the HMF model, the latter goes  to zero as $1/N^{1/9}$ in the  QSS   regime \cite{liap}, while it remains finite at equilibrium at this energy value. 
We  normalized the   data to the corresponding standard   deviations in order to compare them. 
This  comparison clearly shows   that  strong correlations are present   in our case indicating a deviation from  the standard  Central Limit Theorem,  as in \cite{UCC}. A more complete study on this point will be reported   elsewhere.  In any case the  arguments  discussed  in \cite{fulvio2}, and in particular  in Fig. 1(a) of \cite{fulvio2}, do not apply to the  microcanonical  QSS.  The effects become increasingly stronger  when the  initial magnetization gets closer and closer to $1.$   
\\

\emph{(ii) The energy distribution of the QSS cannot be of the Boltzmann-Gibbs exponential form.} \\

As clearly seen in Fig. 1(a) of \cite{fulvio2}, the one-body momentum distribution is {\it non}-Gaussian. This excludes the BG exponential form as the energy distribution in full phase space $\Gamma$, as proved ({\it by reduction to the absurd}) in what follows. This analytical fact can easily be enlarged to $n$-dimensional rotators, $n>1$: $n=1,2,d$, and $\infty$ correspond respectively to the Ising, $XY$, Heisenberg, and spherical models. In what follows we illustrate it,
however, for $n=2$, the case addressed in \cite{fulvio2}. \\

\textbf{Theorem} 
\vskip 0.5cm 
Let $H=K+V$ be the Hamiltonian of a $M$-body classical system, where the kinetic energy is $K=\sum_{i=1}^M \frac{l_i^2}{2I}\;,I>0$, and the potential energy $V$ contains arbitrary {\it integrable} one-body, two-body, three-body,... terms, concerning  (isotropic or anisotropic) rotators localized at a (irregular or regular $d$-dimensional) lattice (each term might be characterized by a distance-dependent coupling constant, which can be summable or not in the $M\to \infty$ limit). And let the one-momentum marginal probability distribution 
$$p(l_1)\equiv \int dl_2\,dl_3...dl_M\,d\theta_1\,d\theta_2\,...d\theta_M \,p(l_1,l_2,...,l_M,\theta_1,\theta_2,...,\theta_M)$$
   associated with the joint distribution 
$$p(l_1,l_2,...,l_M,\theta_1,\theta_2,...,\theta_M)$$ 
be {\it non-Gaussian}. Then 
$$p(l_1,l_2,...,l_M,\theta_1,\theta_2,...,\theta_M)$$
 {\it cannot} be proportional to $$e^{-\beta\,H(l_1,l_2,...,l_M,\theta_1,\theta_2,...,\theta_M)}\,,\,\beta>0 ~~.$$ \\
\vskip 0.25cm
{\bf Proof:}
\vskip 0.15cm
Assume that
$$p(l_1,l_2,...,l_M,\theta_1,\theta_2,...,\theta_M)=\frac{e^{-\beta H}}{\int e^{-\beta H}} ~~,$$
then 
$$p(l_1)=\frac{\int dl_2\,dl_3...dl_M\,d\theta_1\,d\theta_2\,...d\theta_M\,e^{-\beta H}}{\int dl_1\,dl_2\,dl_3...dl_M\,d\theta_1\,d\theta_2\,...d\theta_M\,e^{-\beta H}}=
$$
$$
=\frac{\int dl_2\,dl_3...dl_M\,e^{-\beta K}}{\int dl_1\,dl_2\,dl_3...dl_M\,e^{-\beta K}}
\frac{\int d\theta_1\,d\theta_2\,d\theta_3...d\theta_M\,e^{-\beta V} }
{\int d\theta_1\,d\theta_2\,d\theta_3...d\theta_M\,e^{-\beta V}} = \frac{ e^{-\beta l_1^2/2I \prod_{i=2}^M \int dl_i \, e^{-\beta l_i^2/2I}}}
{\prod_{i=1}^M \int dl_i \,e^{-\beta l_i^2/2I}}=$$
$$=\frac{e^{-\beta l_1^2/2I}}{\int dl_1\,e^{-\beta l_1^2/2I}} ~~,$$
which is a Gaussian, thus contradicting the hypothesis. {\it Q.E.D.}

Hence, the Central Limit Theorem verification exhibited in \cite{fulvio2}
refers to a necessary but insufficient property.\\

\vskip0.5cm
\begin{figure}
\begin{center}
\epsfig{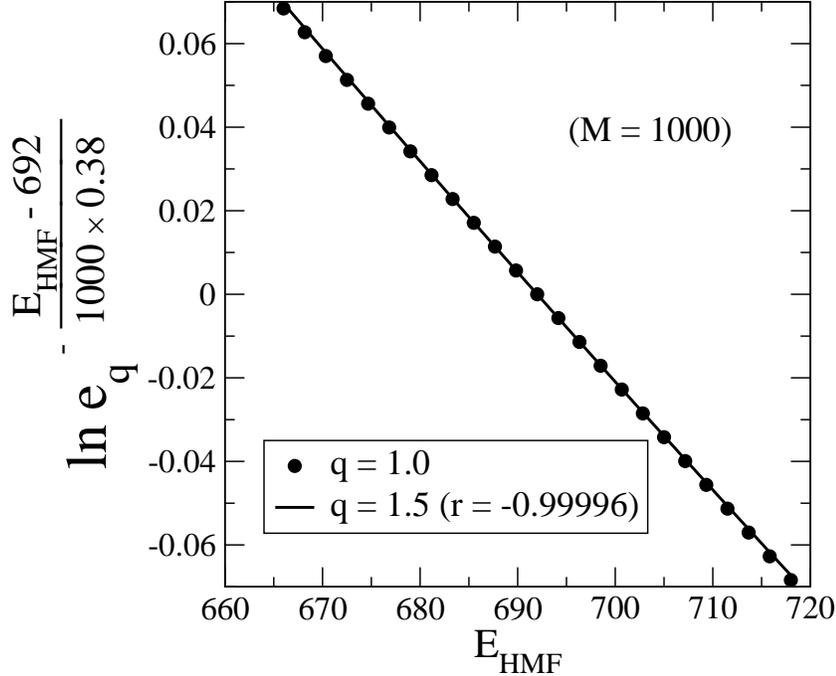}
\end{center}
\caption{
The statistical weights corresponding to BG statistics (full circles), and to a typical illustration of nonextensive statistics ($q=1.5$; solid line), for the $M=1000$ example exhibited in Fig. 3(d) of \cite{fulvio2}, see  text  for details.
}
\end{figure}

\vskip 0.5cm 
\noindent
\emph{(iii) The Baldovin-Orlandini results do not exclude the nonextensive statistical mechanics $q$-exponential form.}\\

Contrary to what is stated in \cite{fulvio2},
$q$-statistics \cite{cost1}
cannot be excluded on the basis of what they exhibit in their Fig. 3(d). Indeed, 
in Fig. 2 we plot 
the statistical weights corresponding to BG statistics (full circles), and to a typical illustration of nonextensive statistics ($q=1.5$; solid line), for the $M=1000$ example exhibited in Fig. 3(d) of \cite{fulvio2}.
We used nearly the same number of points of Fig. 3(d) in the interval $E_{HMF}\in [664,720]$, which covers $4\%$ around its central value $692$. The linear correlation is $r=-0.99996$ ($r=-0.99997$ in \cite{fulvio2} for a similar analysis).
The remarkable closeness of the $q=1$ and $q=1.5$ examples comes from two facts. The first one is that 
 $$e_q^{x}\equiv [1+(1-q)x]^{1/(1-q)}=e^x[1+\frac{1}{2}(q-1)x^2+...], ~~ for ~~x\to 0.$$ 
Notice the absence of the term {\it linear} in $x$ in the latter  expression. While  the  second one is that
 the spreading around the central value $692$ is very small.
Notice also that we rescaled $T= 0.38$ by a factor $M$, which corresponds to multiplying the Hamiltonian (1) of \cite{fulvio2} by M and then rescaling time in the kinetic energy (see \cite{alfaxy}).
This makes the {\it microscopic} two-body interaction of the model {\it independent} from the {\it macroscopic} quantity $M$, as desirable. Consider also that, in the $q$-statistical weight of Fig. 2, we have subtracted from $E_{HMF}$ a characteristic macroscopic energy, namely the  central value $692$. This procedure is long known to be mandatory (see  \cite{mendes}) for any $q \ne 1$ in order to preserve the invariance of probabilities with regard to an uniform shift of the energy scale, i.e., in order to preserve the freedom of the choice of the zero of energies. Such subtraction, although allowed for all values of $q$, is clearly {\it not} mandatory for the BG exponential weight ($q=1$), since the desired invariance is anyhow guaranteed by the fact that the exponential of a sum factorizes into the product of exponentials (whereas the $q$-exponential of a sum does  {\it not} factorize into the product of $q$-exponentials if $q \ne 1$).

\section{Conclusions}

On the  basis  of the  points   here discussed, the   results   presented   in \cite{fulvio2} should be considered  with extreme care, and  by no means they  close the  ongoing debate on the possible application, in appropriate limits, of generalized  $q$-statistics  to long-range  Hamiltonian systems. Finally, let us mention for  completeness that neither the  arguments recently advanced  in \cite{ruffo}  exclude   the  possible application of  $q$-statistics, especially for  initial magnetization close to $1$, but  this point   is beyond the  scope of the present  paper  and   it will be discussed  elsewhere.

\section*{Acknowledgments}

We acknowledge useful discussions with F. Baldovin.
A.P. and A.R. acknowledge
financial support from the PRIN05-MIUR project "Dynamics and
Thermodynamics of Systems with Long-Range Interactions". C.T. acknowledges financial support from the Brazilian Agencies Pronex/MCT, CNPq and Faperj.

\end{document}